\documentclass[aps]{revtex4}
\usepackage[dvips]{graphicx}
\usepackage{amssymb,amsmath,empheq}

\begin{document}

\title{Interacting length scales in the reactive-infiltration instability} 
\author{Piotr Szymczak}
\affiliation{Institute of Theoretical Physics, Faculty of Physics, University of Warsaw,  
Ho\.{z}a 69, 00-618, Warsaw, Poland}
\email{Piotr.Szymczak@fuw.edu.pl}

\author{Anthony J.C. Ladd}
\affiliation{Chemical Engineering Department, University of Florida,
Gainesville, FL  32611-6005, USA}

\begin{abstract}
The reactive-infiltration instability, which develops when a porous matrix is dissolved by a flowing fluid, contains two important length scales. Here we outline a linear stability analysis that simultaneously incorporates both scales. We show that the commonly used ``thin-front'' model is a limiting case of a more general theory, which also includes convection-dominated dissolution as another special case. The wavelength of the instability is bounded from below, and lies in the range $1\mbox{mm}$ to $1\mbox{km}$ for physically reasonable flow rates and reaction rates. We obtain a closed form for the growth rate when the change in porosity is small.
\end{abstract}

\maketitle

\section{Introduction}

\begin{figure}
\center\includegraphics[width=5.0in]{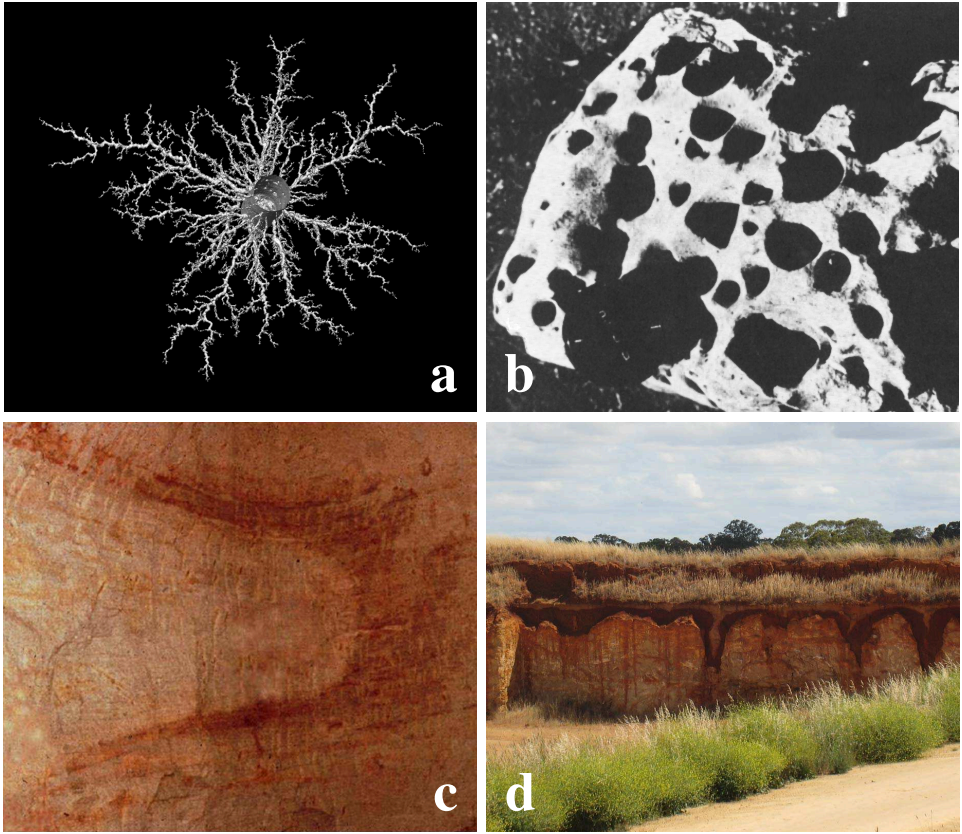}
\caption{Examples of patterns produced by the reactive-infiltration instability: (a) wormholes ($30\mbox{cm}$ long) produced during carbonate acidization~\citep{McDuff2010}, (b) holes formed by limestone dissolution ($5-10 \mbox{cm}$ across) \citep{Ortoleva1987}, (c) a uranium roll ($\sim 1 \mbox{m}$) and (d) terra rossa fingers ($\sim 10 \mbox{m}$). Images are reproduced by permission (see acknowledgments).}\label{fig:examples}
\end{figure}

The reactive-infiltration instability~\citep{Ortoleva1994} is an important mechanism for pattern development in geology, with a range of morphologies (see Fig.~\ref{fig:examples}) and scales, from cave systems running for hundreds of miles~\citep{Ford2007} to laboratory acidization on the scale of centimeters~\citep{Daccord1987}. In this paper we show that the instability is characterized by two length scales; an upstream length where the material is fully dissolved, and a downstream length over which it transitions to the undissolved state. Previous work~\citep{Chadam1986,Ortoleva1987,Sherwood1987,Hinch1990} considered one or the other of these lengths to be dominant, which limits the applicability of their results. In particular we show that the thin-front limit~\citep{Chadam1986,Ortoleva1987} is only valid when a particular combination of reaction rate ($r$), fluid velocity ($v_0$), and diffusion constant ($D$), $D r/v_0^2$ is large. Here we develop a general theory of the reactive-infiltration instability, valid for all flow rates and reaction rates; in addition we obtain closed-form solutions in the limit where the change in permeability is small.

\section{Dissolution of porous media}

When a porous matrix is infiltrated by a reactive fluid, a front develops once all the soluble material at the inlet has dissolved. This front propagates into the matrix as illustrated in the inset to Fig. 2, which shows its position (solid line) at a later time.  A planar front develops perturbations because of the feedback between flow and dissolution \citep{Chadam1986,Sherwood1987,Hinch1990}. Upstream of the front, all the soluble material has dissolved and the porosity is constant, $\phi = \phi_1$. Ahead of the front the porosity gradually decays to its value in the undisturbed matrix, $\phi = \phi_0$.

\begin{figure}
\center\includegraphics[width=5.0in]{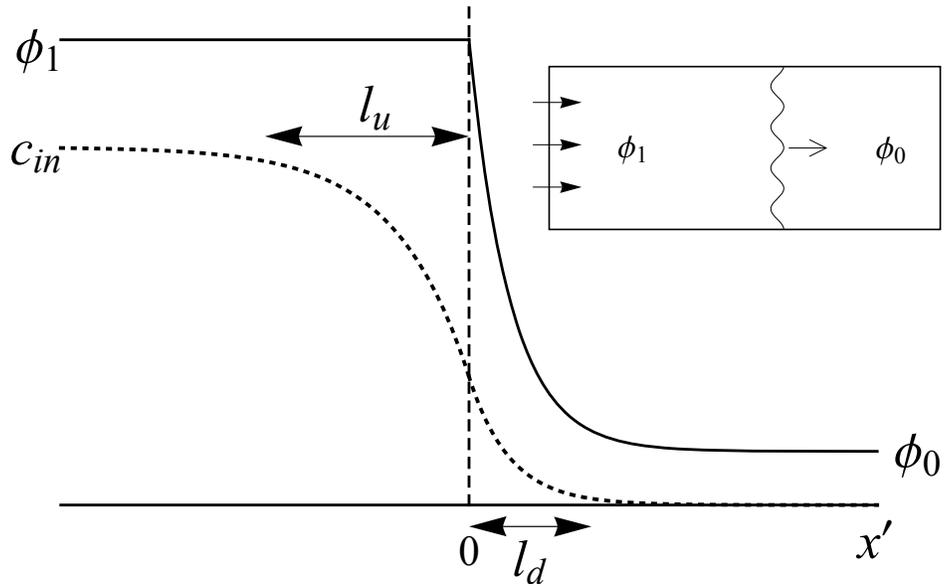}
\caption{Concentration and porosity profiles in the moving front frame $x^\prime = x - Ut$; the position of the front is indicated by the dashed vertical line. The concentration profile decays with different length scales, $l_u$ and $l_d$, in the upstream ($x^\prime < 0$) and downstream ($x^\prime > 0$) regions.}\label{fig:profiles}
\end{figure}

Dissolution of a porous matrix can be modeled by coupled equations describing flow, reactant transport and dissolution:
\begin{align}
\partial_t \phi + \nabla \cdot \mbox{\boldmath$v$} = 0, \ \ \ \ \mbox{\boldmath$v$} &= -K(\phi) \nabla p/\mu, \label{eq:p2D} \\
\partial_t (\phi c) + \mbox{\boldmath$\nabla$} \cdot (\mbox{\boldmath$v$} c) - \mbox{\boldmath$\nabla$} \cdot \mbox{\boldmath$D$} \mbox{\boldmath$\nabla$} c &= - r c \theta(\phi_1 - \phi), \label{eq:c2D} \\
c_{sol} \partial_t \phi &= r c \theta(\phi_1 - \phi), \label{eq:e2D}
\end{align}
where $\theta$ is the Heaviside step function, $\mbox{\boldmath$v$}$ is the superficial velocity, and $c_{sol}$ is the concentration of soluble material. In the upstream region, where all the soluble material has dissolved ($\phi=\phi_1$), the reaction terms vanish and the transport equation reduces to a convection-diffusion equation. We assume a constant diffusivity $\mbox{\boldmath$D$} = D \mbox{\boldmath$1$}$ and reaction rate $r$, which captures the essential characteristics of the reactive-infiltration instability. A complete description of solute dispersion and reaction in porous media is complex \citep{Golfier2002,Panga2005}, but  more general dispersion $\mbox{\boldmath$D$}(\phi, \mbox{\boldmath$v$})$ and reaction $r(\phi)$ coefficients can be incorporated within the same framework.

The rate of increase in porosity ($\partial_t \phi$) is much smaller than the reaction rate ($r$) because the molar concentration of solid ($c_{sol}$) is typically orders of magnitude larger than the reactant concentration ($c$). This time scale separation can be quantified by introducing  the acid capacity number, $\gamma_a = c_{in}/[c_{sol}(\phi_1-\phi_0)]$, which corresponds to the volume of porous matrix dissolved by a unit volume of reactant. In typical geophysical systems the reactant is dilute $c_{in} \ll c_{sol}$ and therefore  $\gamma_a \ll 1$; for example, when calcite is dissolved by aqueous $\rm CO_2$, $\gamma_a\sim 10^{-4}$. This allows us to drop the time derivatives in the flow and transport equations (\ref{eq:p2D}--\ref{eq:c2D}) and treat the velocity and concentration fields as stationary, slaved to the porosity field by means of the erosion equation \eqref{eq:e2D}.

The resulting equations are closed by auxiliary conditions far from the front:
\begin{align}
\mbox{\boldmath$v$}(-\infty) &= v_0\mbox{\boldmath$e$}_x & v_y(\infty) &= 0, \\
c(-\infty) &= c_{in}, & c(\infty) &= 0, \\
\phi(-\infty) &= \phi_1, & \phi(\infty) &= \phi_0,
\end{align}
where $\mbox{\boldmath$e$}_x$ is a unit vector in the direction of the flow ($x$). Far upstream ($x \rightarrow -\infty$) the fluid velocity ($v_0$) and reactant concentration ($c_{in}$) are uniform, and the matrix is fully dissolved. Far downstream ($x \rightarrow \infty$) the reactant has been entirely consumed, the matrix is undisturbed and the pressure is uniform across the sample ($\partial_y p = 0$).

Equations \eqref{eq:p2D}--\eqref{eq:e2D} have steady one-dimensional solutions, $c_b(x^\prime)$ and $\phi_b(x^\prime)$, in a frame $x^\prime = x - Ut$ moving with a constant velocity $U$:
\begin{align}
v_0\partial_{x^\prime} c_b - D \partial_{x^\prime}^2 c_b &= - r c_b \theta(\phi_1 - \phi_b), \label{eq:c1D} \\
-c_{sol} U \partial_{x^\prime} \phi_b &= r c_b \theta(\phi_1 - \phi_b). \label{eq:e1D}
\end{align}
The fluid velocity $v_0$ is constant throughout the domain  (in one dimension), and a mass balance on the reactant consumption requires that $U = \gamma_a v_0$. By taking the limit of small acid capacity ($\gamma_a \rightarrow 0$) the term proportional to $U$ can be dropped from Eq.~\eqref{eq:c1D}, but not from \eqref{eq:e1D} where it sets the dissolution time scale $t_d = l_d/U$.

Upstream of the front, indicated by the dashed line in Fig.~\ref{fig:profiles}, all the soluble material has  dissolved and the porosity is constant, $\phi = \phi_1$. Nevertheless, because of the diffusive flux, the concentration profile is not uniform in this region, but decays from its inlet value $c_{in}$ over a characteristic length $l_u = D/v_0$. Convection-dominated theories \citep{Sherwood1987,Hinch1990} neglect this scale, setting the concentration at the front to $c_{in}$, but our analysis shows that this is a singular limit (see Eq.~\ref{eq:w1lim}) and that even small diffusive contributions to the reactant flux make a large difference to the growth rate.

Downstream from the front, material is still being dissolved and here the concentration decays with a different length scale, $l_d = 2D/(\sqrt{v_0^2 + 4Dr} - v_0)$; the porosity returns to its initial value $\phi_0$ on the same scale. If the thickness of the downstream front is neglected then the reactive-infiltration instability can be mapped to a thin-front problem \citep{Chadam1986,Ortoleva1987}, with an $r$-independent growth rate (see Eq.~\ref{eq:w1diff}). These results have been widely used to draw inferences about the mechanisms and growth rates for morphological changes in rocks, but the range of validity of this limit is more restricted than is generally realized.

\citet{Chadam1986} and \citet{Ortoleva1987} proposed that the front-thickness can be neglected whenever the acid capacity is small ($\gamma_a \ll 1$), a condition that is widely applicable in nature and which is implicit in Eq.~\eqref{eq:c1D}. By including the acid capacity in their definition of reaction rate, $r^\star = r/\gamma_a$, they necessarily take the fast-reaction limit $r\rightarrow\infty$ as $\gamma_a\rightarrow0$ (keeping $r^\star$ finite) \citep{Ortoleva1987a}. Missing from this analysis is what physical quantity $r$ must be large with respect to; Eq.~\eqref{eq:c1D} suggests that a sharp front, meaning $l_d \ll l_u$ (Fig.~\ref{fig:profiles}), will only occur when $r \gg v_0^2/D$. Thus there is no general reduction of the reactive-infiltration instability to a thin-front problem; it is only appropriate as a limiting case when the dimensionless combination of reaction rate, flow rate, and diffusion, $H = Dr/v_0^2  \gg 1$.

The importance of the interplay between reaction, diffusion and convection in a dissolving rock matrix was first noted by \citet{Lichtner1988} and \citet{Phillips1990}, while the relevance of the parameter $H$ to wormhole growth was recognized by~\citet{Steefel1990}; nevertheless, they did not incorporate their insights into a stability analysis. However, \citet{Aharonov1995} discussed a similar interaction of length scales in the related problem of melt flow in the mantle; here the interplay of matrix compaction and solubility gradient leads to an instability in an otherwise steady, non-propagating porosity profile. Fracture dissolution is also characterized by an instability in a non-propagating (but time-dependent) front~\citep{Szymczak2012}. However, the impact of diffusion in systems with propagating versus non-propagating fronts is fundamentally different: in moving front problems, diffusion can completely stabilize the growth \citep{Chadam1986}, whereas in the case of non-propagating profiles it weakens the growth but does not make it stable \citep{Aharonov1995,Szymczak2011}. In this paper we present a new analysis of the instability in a steadily propagating dissolution front, which includes both upstream (where the material is fully dissolved) and downstream regions; we recover previous results \citep{Chadam1986,Sherwood1987,Hinch1990} as limiting cases.

\section{Linear stability analysis}

Stationary one-dimensional solutions of Eq.~\eqref{eq:p2D} form the base state for the linear stability analysis:
\begin{align}
\frac{c_b}{c_{in}} &= 1 - \frac{e^{x^\prime/l_u}}{1+Pe}, & &\phi_b = \phi_1 && x^\prime < 0; \label{eq:busol} \\
\frac{c_b}{c_{in}} &= \frac{e^{-x^\prime/l_d}}{1+Pe^{-1}}, & &\frac{\phi_b-\phi_0}{\phi_1-\phi_0} = e^{-x^\prime/l_d} && x^\prime > 0. \label{eq:bdsol} 
\end{align}
The P\'eclet number is defined on the scale of the downstream length, $Pe = v_0 l_d/D$, and is equal to the ratio of upstream and downstream length scales, $l_d/l_u$; it is a function of the dimensionless parameter $H=Dr/v_0^2$; $Pe = 2/(\sqrt{1+4H}-1)$. Although $Pe$ is usually based on pore size or sample size, geophysical systems are typically unbounded; then the reactant penetration length is the largest and most important length scale.

Perturbations to the porosity, velocity and concentration fields are determined by linearizing about the base state; {{\it e.g. }}
\begin{equation}
\phi(x^\prime,y,t) =\phi_b(x^\prime) + \delta \phi(x^\prime) \sin(uy) e^{\omega t}. \label{eq:dp}
\end{equation}
The result of the stability analysis is a fifth-order differential equation for the downstream porosity perturbation $\delta \phi$, with solutions that depend on $Pe$, $u$, $\omega$, and $K(\phi)$. Here we take a $\phi^3$ relation for the permeability
\begin{equation}\label{eq:perm}
K(\phi) = K_0\left(\frac{\phi}{\phi_0}\right)^3,
\end{equation}
and solve for the downstream  $\delta \phi$ using a spectral method~\citep{Boyd1987}. Boundary conditions at the front were constructed by matching to analytic solutions for the upstream perturbations in velocity and concentration. Two boundary conditions suffice to determine $\delta \phi$ for a given $\omega$, and the remaining boundary condition is used to determine the growth rate $\omega(u, Pe, \Delta)$, where
\begin{equation}
\label{eq:Delta}
\Delta = (\phi_1-\phi_0)/\phi_0
\end{equation}
is the porosity contrast. Details of the linear stability analysis are given in the Auxiliary Material.

\begin{figure}
\center\includegraphics[width=5.0in]{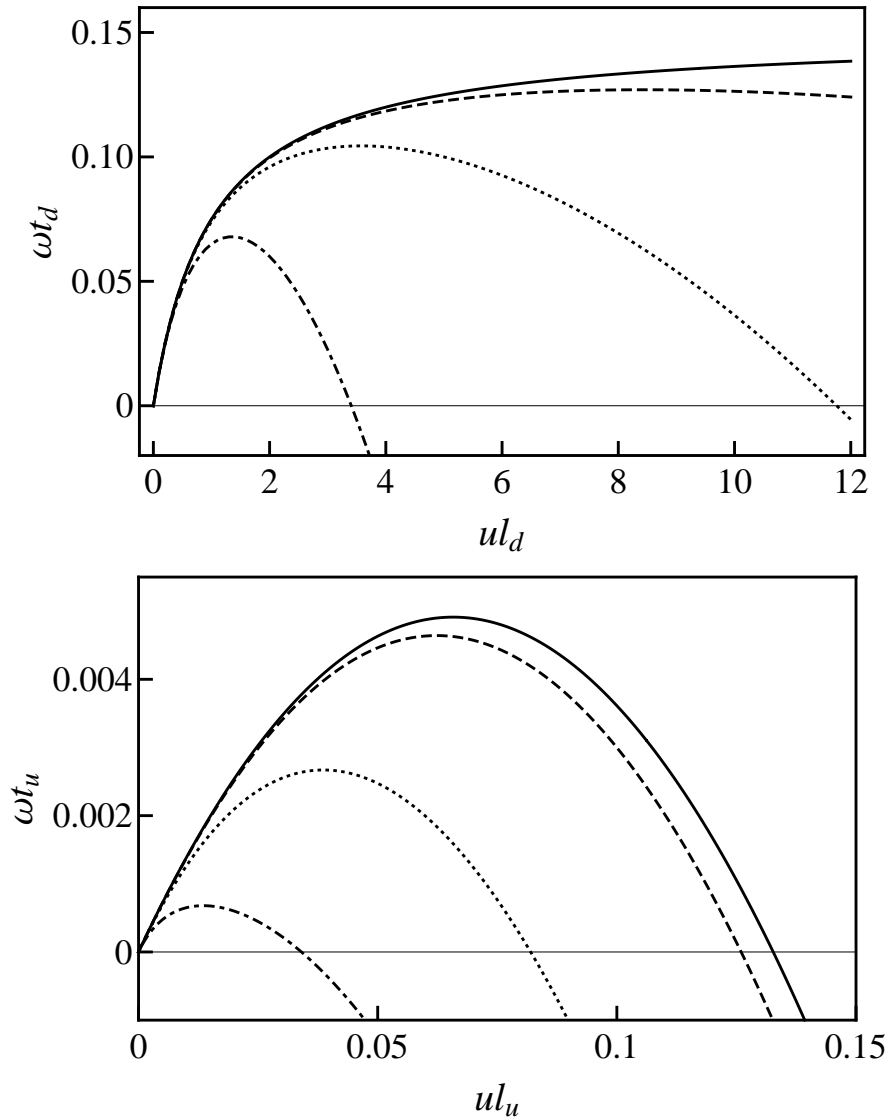}
\caption{Instability growth rates for a small porosity contrast, $\Delta= 0.1$. The upper panel shows the dispersion curves in the downstream scaling for large values of P\'eclet number: $Pe = \infty$ (solid), $Pe=10^4$ (dashed), $Pe=10^3$ (dotted) and $Pe=10^2$ (dot-dashed). The lower panel shows the dispersion curves in the upstream scaling for small values of P\'eclet number: $Pe = 0$ (solid), $Pe=1$ (dashed), $Pe=10$ (dotted) and $Pe=100$ (dot-dashed) respectively}\label{fig:W0}
\end{figure}

The connection between the various limiting cases $Pe \gg 1$, $Pe \sim 1$, and $Pe \ll 1$ can be made explicit by developing a perturbation expansion in the porosity contrast $\Delta$. The final result for the growth rate is
\begin{equation}
\omega t_d =  \frac{1}{2}\left(Pe-\sqrt{Pe^2+4 u^2 l_d^2}\right) + \Delta\, \omega_1 t_d + O(\Delta^2), \label{eq:W0}
\end{equation}
where the time scale $t_d = l_d/\gamma_a v_0$, and $\omega_1(ul_d, Pe)$ is a simple but lengthy algebraic function (see Auxiliary Material). Characteristic dispersion curves for convection-dominated dissolution ($Pe \gg 1$) are shown in the upper panel of Fig.~\ref{fig:W0} for a small porosity contrast $\Delta = 0.1$. For large P\'eclet numbers, Eq.~\eqref{eq:W0} can be replaced by a simpler expression,
\begin{equation}
\omega t_d =  \frac{3\Delta\, u l_d}{2(1+u l_d)}- \frac{u^2 l_d^2}{Pe} + O(Pe^{-2}), \label{eq:w1lim}
\end{equation}
with results that are indistinguishable from Eq. \eqref{eq:W0} on the scale of the upper panel in Fig.~\ref{fig:W0}.

In the convective limit ($Pe \rightarrow \infty $), $\omega$ rises monotonically with increasing wavevector, reaching an asymptotic value $\omega t_d  = 3\Delta/2$ as in \citet{Hinch1990}. However, even a small diffusivity cuts off the short wavelengths ($u l_d > 1$), leading to a pronounced maximum in the growth rate. This implies that there will be a strong wavelength selection even in highly convective flows and that short wavelength perturbations will not grow. Thus, the convective limit is singular; unstable terms saturate for large values of $ul_d$ and are eventually overwhelmed by the diffusive stabilization.

As the P\'eclet number decreases, diffusional stabilization reduces the growth rate and pushes the range of unstable wavelengths towards $u = 0$. When $Pe \ll 1$, the dispersion relation \eqref{eq:W0} takes a particularly simple form in variables scaled by the upstream length and time, $u l_u$ and $\omega t_u$, where the upstream time scale $t_u = l_u/\gamma_a v_0$. Physically, this rescaling is associated with the change in length scale from convection-dominated infiltration, $l_d = v_0/r \gg l_u$, to diffusion-dominated infiltration, $l_d \ll l_u = D/v_0$. The lower panel of Fig.~\ref{fig:W0} shows that the dispersion relation in this scaling reaches a limiting form for small $Pe$,
\begin{equation}
\omega t_u = \frac{3\Delta}{2}u l_u + \left(\frac{1}{2} + \frac{3\Delta}{4}\right)\left(1 - \sqrt{1+4u^2 l_u^2}\right) + O(Pe^2), \label{eq:w1diff}
\end{equation}
shown by the solid line; this is the thin-front limit ($l_d / l_u \rightarrow 0$) considered in~\citet{Ortoleva1987}, with $\Delta \ll 1$. 

The most important result from an analysis of the reactive-infiltration instability is the wavelength of the fastest growing mode, $\lambda_{max}(Pe,\Delta) = 2\pi/u_{max}$. Natural patterns are expected to develop on this scale \citep{Ortoleva1987}, at least initially, since this particular wavelength grows exponentially faster than neighboring ones, with a time scale $t_{max} = 1/\omega_{max}=1/\omega(\lambda_{max})$. The fastest-growing length and time scales are shown in Fig.~\ref{fig:umax} over a range of Darcy velocities, from $10^{-8} - 10^{-1}\mbox{cm}\mbox{\,s}^{-1}$. The lower end of the scale covers the range of naturally occurring flow rates, while much higher velocities (up to $0.1\,\mbox{cm}\mbox{\,s}^{-1}$) are found in reservoir acidization. Results are shown for three different reaction time scales: $r^{-1} = 10^8 \mbox{s}$, which is characteristic of slowly dissolving minerals such as quartz and certain redox reactions, $r^{-1} = 10^4 \mbox{s}$, which is typical of the dissolution of clays, and $r^{-1} = 1\mbox{s}$ which is characteristic of fast dissolving minerals such as calcite or gypsum; we will consider some specific examples in the Discussion. Throughout we will assume a constant diffusion coefficient, $D=10^{-5}\mbox{cm}^2\mbox{\,s}^{-1}$.

Figure~\ref{fig:umax} shows there is a lower bound to the wavelength and time scale with respect to the flow velocity, which occurs around $Pe = 10$ in each case. Starting in the thin-front limit ($v_0 \sim 10^{-8} \mbox{cm}\mbox{\,s}^{-1}$), an increasing velocity reduces the wavelength by decreasing the upstream penetration length $l_u = D/v$; in this region ($Pe < 1$) the downstream penetration length is small. However, as $v_0$ increases further the downstream length begins to grow and when $Pe > 1$ takes over as the dominant scale; in this case an increase in velocity increases the penetration length and so the scale of the instability grows. Interestingly the plot suggests that instability wavelengths will fall in a range between millimeters and a few hundred meters, since the highest flow rates are typically associated with reservoir acidization, where reaction rates are high (red curves). Similarly, there is a lower bound to the time scale; once the downstream penetration length starts to take over the growth rate of the instability becomes independent of velocity. It should be noted that in order to obtain universal curves we have plotted $\gamma_a t_{max}$ in Fig.~\ref{fig:umax}; the dissolution time for a particular mineral can be found by dividing by the acid capacity (typically $\gamma_a \sim 10^{-6}-10^{-4}$).

The most unstable wavelength $\lambda_{max}$ decreases with increasing porosity contrast up to $\Delta \sim 1$. In the diffusive regime it then saturates but in the convective regime $\lambda_{max}$ increases sharply with velocity, as shown by the solid squares in the upper panel of Fig.~\ref{fig:umax}. This results from the appearance of a long wavelength maximum in the dispersion curve at high porosity contrasts ($\Delta \sim 10$), which persists from $Pe \approx 10$ to the convective limit \citep{Szymczak2011a}. The analytic theory \eqref{eq:W0} remains valid up to $\Delta =1$ (dashed lines) as can be seen by the comparison with numerical results (solid circles).

\section{Discussion}

A reactive-infiltration instability can occur in almost any system in which chemical dissolution is coupled with fluid flow. Variations in reaction rate ($r$) and flow rate ($v_0$) give rise to a wide range of length scales, from centimeter scale redox fronts in siltstones~\citep{Ortoleva1994} to kilometer-long scalloping of a dolomitization front~\citep{Merino2011}. The span of timescales is also large; acidized plaster~\citep{Daccord1987} and limestone cores~\citep{Hoefner1988} or salt flushed with water~\citep{Kelemen1995,Golfier2002} finger in minutes, while geological structures evolve over hundreds of thousands of years. Groundwater velocities are usually small, $v_0 \approx 10^{-8} - 10^{-5} \mbox{cm}\mbox{\,s}^{-1}$, while the timescale for dissolution ($r^{-1}$) varies from seconds to years; thus both diffusion-dominated ($Dr/v_0^2 \gg 1$) and convection-dominated ($Dr/v_0^2 \ll 1$) dissolution can occur. 

\begin{figure}
\center\includegraphics[width=5.0in]{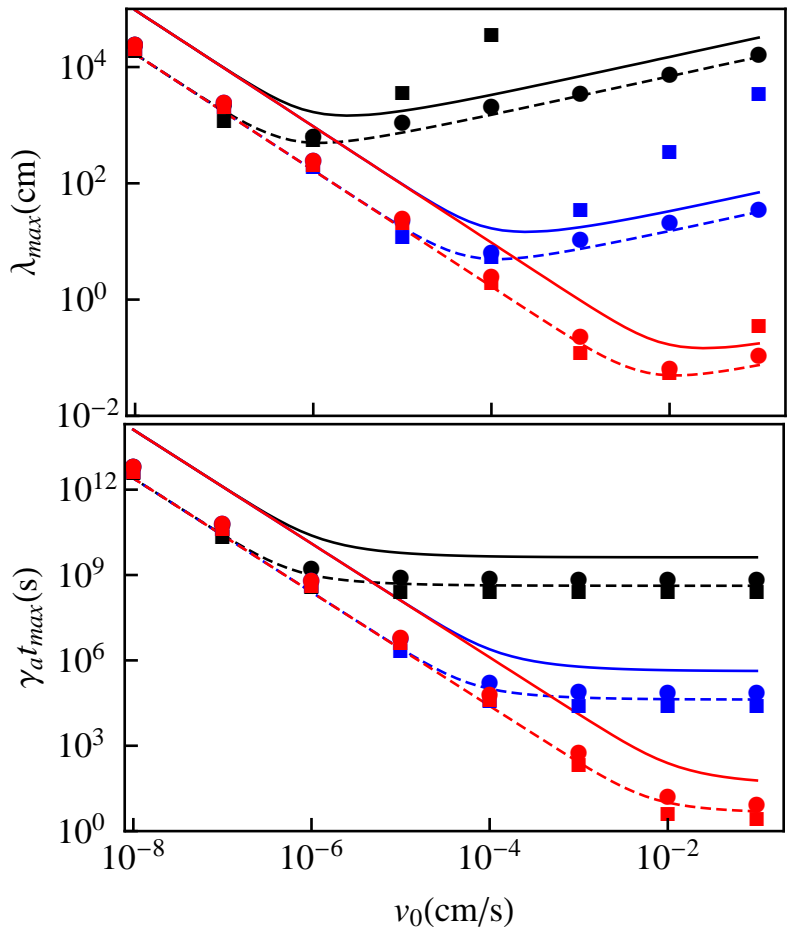}
\caption{The maximally unstable wavelength, $\lambda_{max} = 2\pi/u_{max}$ (upper panel), and time scale, $t_{max} =1/\omega_{max}$ (lower panel), are shown for different reaction rates: $r=10^{-8}\mbox{\,s}^{-1}$ (black), $r=10^{-4}\mbox{\,s}^{-1}$ (blue), and $r=1\mbox{\,s}^{-1}$ (red). For each reaction rate results are shown at three different porosity contrasts: $\Delta = 0.1$ (solid line), $\Delta = 1$ (dashed line and circles), and $\Delta=10$ (squares). Analytic results from \eqref{eq:W0} are shown by lines and numerical results from the spectral code are shown by the solid symbols. The acid capacity $\gamma_a$ is needed to determine the actual dissolution time scale for a specific mineral.}\label{fig:umax}
\end{figure}

The formation of salt sinkholes is an example of diffusion-dominated dissolution; here $r \approx 2 \times 10^{-4} \mbox{\,s}^{-1}$ and $v_0$ is in the range $3 \times 10^{-7}-3 \times 10^{-6} \mbox{cm}\mbox{\,s}^{-1}$~\citep{Shalev2006}, which means it is in the diffusive regime ($H > 1$). For the large porosity contrast typical of salt dissolution ($\Delta \approx 10$), the maximum unstable wavelength (Fig.~\ref{fig:umax}) is then $\lambda_{max} \approx 0.7-7\mbox{m}$, which is in the range of results reported in~\citet{Shalev2006}. The associated timescales are of the order of $1.5-150$ years (with $\gamma_a \approx 0.18$); thus sinkholes would be expected to develop over tens of years, which is comparable with experimental observations~\citep{Shalev2006}.

Even though the natural flow rates in most rock formations are small, convection-dominated dissolution can still occur, particularly if only a small fraction of the grains are reactive. Relevant examples include uranium deposition, in which a solution containing soluble uranium salts reacts with pyrites (FeS) embedded in a sandstone matrix; this alters the redox potential and causes uranium to precipitate as uraninite (uranium oxide) \citep{Lake2002}. Since the pyrites constitute only about 2\% of the total rock matrix~\citep{Dewynne1993}, the rate of uranium precipitation is small, $r \approx 10^{-8}\mbox{\,s}^{-1}$~\citep{Lichtner1992}. Oxidation of pyrites produces sulfuric acid, which dissolves some of the rock matrix; the increase in porosity ($\Delta \approx 4$, \cite{Dewynne1993}) causes a reactive-infiltration instability in the uraninite front. Within the typical range of sandstone permeability, dissolution can be either convection-dominated or diffusion-dominated. The thin-front limit applies when the flow velocity is less than $10^{-7}\mbox{cm}\mbox{\,s}^{-1}$; here the scale of the front is predicted to be about $20\mbox{m}$, varying as $v_0^{-1}$. However, at higher flow velocities the scale decreases more slowly because of the transition to convection-dominated dissolution, with a minimum wavelength of about $6\mbox{m}$ (at $v_0 = 10^{-6} \mbox{cm}\mbox{\,s}^{-1}$). At $v_0 = 10^{-5} \mbox{cm}\mbox{\,s}^{-1}$ the scale is again about $20\mbox{m}$, whereas the thin-front prediction would be more than an order of magnitude smaller. Field observations indicate spacings between uraninite fingers in the range $1 \mbox{m}$ to $1 \mbox{km}$~\citep{Dahlkamp2009}; in the latter case, since the fingers are very prominent, the observed spacing most likely reflects a non-linear selection process, which eliminates many smaller channels (see below).

Fluid velocities during acidization of carbonate reservoirs are larger than in naturally occurring groundwater flows, $v_0 \approx 10^{-3} \mbox{cm}\mbox{\,s}^{-1} - 0.1 \mbox{cm}\mbox{\,s}^{-1}$~\citep{Economides2000}, and it is frequently assumed that acidization is convection dominated \citep{Sherwood1987,Hoefner1988}. However, rapid dissolution of calcite by concentrated hydrochloric acid (frequently used in acidization), combined with the large reactive surface area of calcite, can lead to reaction rates as high as $10\mbox{\,s}^{-1}$~\citep{Cohen2008a}, although weaker acids, such has acetic or formic acid, have dissolution rates that are $100$ times slower. Thus acidization spans the range from convection-dominated ($H \sim 10^{-4}$) to diffusion-dominated dissolution ($H \sim 100$), but the predicted length scales are always in the sub-centimeter range. At the smallest flow rates ($v_0 \approx 10^{-3} \mbox{cm}\mbox{\,s}^{-1}$), dissolution tends to be diffusion dominated, with a typical length scale of the order of $\lambda_{max} \approx 0.1\mbox{cm}$, independent of reaction rate. At the highest flow rates ($v_0 \approx 0.1\mbox{cm}\mbox{\,s}^{-1}$) the reactant penetrates downstream, leading to significantly larger scales ($\sim 0.1\mbox{cm}$) than would be predicted from the thin-front limit ($\sim 0.001 \mbox{cm}$). Intriguingly there may be a connection between the minimum scale of the instability (Fig.~\ref{fig:umax}) and the optimization of reactant consumption during acidization \citep{Fredd1998,Golfier2002,Panga2005}. Although the dynamics of wormhole formation are highly nonlinear, the interaction of convective, reactive, and diffusive length scales may be similar.

In this work we have presented the results of an analysis of the reactive-infiltration instability, which takes account of the two lengths scales that characterize the concentration field; previous analysis  \citep{Chadam1986,Ortoleva1987,Sherwood1987,Hinch1990,Szymczak2011a} assumed that one length or the other was dominant, corresponding to the limiting cases $H \rightarrow \infty$ \citep{Chadam1986,Ortoleva1987} or $H \rightarrow 0$ \citep{Sherwood1987,Hinch1990,Szymczak2011a}. We have explained the connection between these apparently disparate theories and determined their range of validity; in particular, we have shown that the thin-front limit only holds when $H \gg 1$ and the convective limit when $H \ll 1$. We have given examples of reactive infiltration systems -- in nature, laboratory experiments, and engineered systems -- which span a wide range of $H$, from $H \sim 10^{-4}$ to $H \sim 10^3$.

Finally, we note that there are significant limitations to the use of a linear stability analysis to interpret geological morphologies. First, reaction rates at the Darcy scale are not well understood; field measurements are frequently orders of magnitude smaller than those inferred from laboratory experiments. Moreover, as the instability develops non-linear couplings lead to a coarsening of the pattern, with an increasing wavelength between the fingers~\citep{Szymczak2006}. The competition between different fingers causes the shorter ones to be arrested, which can be clearly seen in the terra-rossa fingers in Fig.~\ref{fig:examples} and also in maps of uraninite formations~\citep{Dahlkamp2009}.

\section*{Acknowledgments}
This work was supported by the US Department of Energy, Chemical Sciences, Geosciences and Biosciences Division, Office of Basic Energy Sciences (DE-FG02-98ER14853). We thank the following individuals for permission to reproduce the images in Fig.~\ref{fig:examples}: (a) Darren McDuff (ExxonMobil Upstream Research) (b) Enrique Merino (University of Indiana) (c) Dr. Robert Gregory (Wyoming State Geological Survey); photograph by C. L. Van Alstine (U. S. Atomic Energy Commission)(d) Les Sampson (Claremont Wines, South Australia).

\bibliographystyle{unsrt}

\section*{Auxilliary material: derivation of instability growth rates}

\newcommand{\Hp}			{\hat {\phi}}
\newcommand{\Hc}			{\hat{c}}
\newcommand{\Hu}			{\hat{u}}
\newcommand{\Hv}			{\hat{v}}
\newcommand{\Ho}			{\hat{\omega}}
\newcommand{\Pek}			{Pe}
\newcommand{\Dx}			{\partial_\xi}
\newcommand{\BB}[1]		{\mbox{\boldmath${#1}$}}
\newcommand{\Vv}		{{\BB{v}}}

The derivation of the growth rate (14) proceeds in four steps; first we derive linearized equations for small perturbations about the downstream and upstream base states. The upstream perturbations can be found in closed form, whereas the downstream perturbations are solutions of a fifth-order differential equation. Next, we use the upstream solutions to create the boundary conditions for the downstream solutions. In the general case the growth rate must then be found numerically. However, for small variations in permeability we can obtain a closed-form solution for the growth rate over the whole range of flow rates and reaction rates.

\subsection{Downstream perturbations}

We begin with a dimensionless form of Eqs.~(1)-(3), scaling length with the downstream penetration length, $\xi = x^\prime/l_d$, $\eta = y/l_d$, $\tau = \gamma_a v_0 t/l_d$, and defining the dimensionless fields $\Hp = (\phi-\phi_0)/(\phi_1-\phi_0)$, $\Hc = c/c_{in}$, ${\hat \Vv} = \Vv/v_0$. We transform to a frame $x^\prime = x - Ut$, moving with the velocity of the steadily propagating front, $U=\gamma_a v_0$ (8):
\begin{align}
\Dx \Hv_\xi + \partial_\eta \Hv_\eta &= 0, \label{cont} \\
\partial_\eta \Hv_\xi -  W \Hv_\xi \partial_\eta \Hp  &= \Dx \Hv_\eta -  W \Hv_\eta \Dx \Hp, \label{compat} \\
\Hv_\xi \Dx \Hc + \Hv_\eta \partial_\eta \Hc  - \Pek^{-1} (\Dx^2 \Hc + \partial_\eta^2 \Hc) &=  - (1+\Pek^{-1})\Hc,  
\label{transport} \\
\partial_\tau\Hp - \Dx \Hp &= (1+\Pek^{-1}) \Hc, \label{erosion} 
\end{align}
where Darcy's equation has been replaced with the more convenient compatibility equation \eqref{compat} by eliminating the pressure. The permeability gradient $W$ is defined by $W(\Hp) = d \ln K /d \Hp$; from Eq. (12) $W(\Hp) = 3\Delta(1+\Delta \Hp)^{-1}$, where $\Delta = (\phi_1 - \phi_0)/\phi_0$ is the porosity contrast (13). Equations~\eqref{cont} and \eqref{transport} have been simplified by dropping terms proportional to the acid capacity number, $\gamma_a$, which is assumed to be small $\gamma_a = c_{in}/[c_{sol}(\phi_1-\phi_0)] \ll 1$. This enforces a time-scale separation between dissolution of the solid and development of the concentration field, which is at steady state on the dissolution time scale. In the downstream scaling, the dimensionless reaction rate $r l_d/v_0 = 1+\Pek^{-1}$. 

The one-dimensional equations for the base profile follow from neglecting the $\tau$ and $\eta$ dependencies:
\begin{align}
\Hv_b &= 1, \\
\Dx \Hc_b - \Pek^{-1} \Dx^2 \Hc &= - (1+\Pek^{-1})\Hc, \label{cbase} \\
- \Dx \Hp &= (1+\Pek^{-1}) \Hc,
\end{align}
which are the dimensionless versions of Eqs. (7) and (8). The downstream ($\xi\rightarrow\infty$) boundary conditions (4)--(6) require decaying solutions for $\Hc_b$ and $\Hp_b$, which for the dimensionless fields are simply:
\begin{align}
\Hc_b = \Hc_b(0)e^{-\xi}, \label{cbased} \\
\Hp_b = e^{-\xi}.
\end{align}
Upstream of the front ($\xi < 0$) the scaled porosity field is always unity while the concentration field must match to the upstream solution, which is yet to be determined.

Beginning with the base solutions for the porosity, velocity, and concentration, we consider infinitesimal perturbations to each field:
\begin{align}
\Hp &= \Hp_b + \delta \Hp \label{hp}\\
\Hc &= \Hc_b + \delta \Hc \label{hc}\\
\Hv_\xi &= 1 + \delta \Hv_\xi, \label{hv} \\
\Hv_\eta &= \delta \Hv_\eta \label{hvperp}
\end{align}
The following linearized equations for the perturbations in aperture, concentration and flow fields are obtained:
\begin{align}
\Dx^2 \delta \hat{v}_\xi + \partial_\eta^2 \delta \hat{v}_\xi &= W\left[(\Dx \Hp_b)(\Dx \delta \hat{v}_\xi) + \partial_\eta^2 \delta \Hp \right], \label{compat1} \\
\delta \hat{v}_\xi \Dx \Hc_b + \Dx \delta \Hc  - \Pek^{-1} (\Dx^2 \delta \Hc + \partial_\eta^2 \delta \Hc) &= - (\partial_\tau-\Dx) \delta \Hp, \label{transport1} \\
(\partial_\tau-\Dx) \delta \Hp &=  (1+\Pek^{-1})\delta \Hc, \label{erosion1}
\end{align}
The incompressibility condition \eqref{cont} was used to eliminate $\delta v_\eta$ from the compatibility relation \eqref{compat}, and the erosion equation \eqref{erosion} was used to replace the reaction term in \eqref{transport}.

Equations \eqref{compat1}--\eqref{erosion1} are linear in the perturbations, and we assume that solutions are normal modes, sinusoidal in $\eta$ and exponential in $\tau$:
\begin{align}
\delta \Hp &= f_{\phi}(\xi)\cos(\Hu\eta)e^{\Ho \tau}, \label{eq:dphi} \\
\delta \Hc &= f_c(\xi)\cos(\Hu\eta)e^{\Ho \tau}, \label{eq:dc} \\
\delta \hat{v}_\xi &= f_v(\xi)\cos(\Hu\eta)e^{\Ho \tau}. \label{eq:dv}
\end{align}
Substituting these expansions into Eqs.~\eqref{compat1}--\eqref{erosion1} leads to coupled equations for the one-dimensional fields $f_\phi(\xi)$, $f_c(\xi)$, and $f_v(\xi)$:
\begin{align}
W\Hu^2 f_{\phi} &= (-\partial_{\xi}^2 + W (\Dx \Hp_b)\Dx + \Hu^2) f_v, \label{cont2} \\
(\Dx \Hc_b) f_v &= (\Dx -\Ho) f_{\phi} + \left[\Pek^{-1} (\partial_{\xi}^2 -\Hu^2) - \Dx\right] f_c, \label{transport2}\\
(1+\Pek^{-1}) f_c &= (-\Dx + \Ho) f_{\phi}. \label{erosion2}
\end{align}
Equations \eqref{cont2}--\eqref{erosion2} can be combined into a fifth-order eigenvalue equation for the evolution of the downstream porosity field,
\begin{equation}\label{eq:pt}
\left[\partial_{\xi}^2 + W e^{-\xi}\Dx - \Hu^2\right]e^\xi {\cal L}_\lambda (\Dx -\Ho) f_{\phi} = W \Hu^2 f_{\phi},
\end{equation}
where the convection-diffusion operator ${\cal L}_\lambda$ is given by
\begin{equation}\label{eq:Hlambda}
{\cal L}_\lambda = \Dx + 1 -\Pek^{-1}(\Dx^2 - u^2 -1).
\end{equation}

In the far-field ($\xi \rightarrow \infty$), only the leading order terms in $e^\xi$ remain,
\begin{equation}\label{eq:ptFF}
(\Dx^2 - \Hu^2)e^\xi{\cal L}_\lambda(\Dx - \Ho) f_{\phi} = 0.
\end{equation}
Then, of the five solutions for $f_\phi$, only two are decaying for all values of $\hat u$,
\begin{equation}
f_\phi(\xi \rightarrow \infty) = A_\phi e^{-(1+\Hu) \xi} + B_\phi e^{\lambda \xi}, \label{eq:ptffsol}
\end{equation}
where $\lambda$ is the negative eigenvalue of the operator ${\cal L}_\lambda$. Thus the three downstream boundary conditions (4)--(6) eliminate three solutions for $f_\phi$.

\subsection{Upstream perturbations}

There is a universal solution for the upstream velocity and concentration fields (denoted by a superscript $u$), which can be used to derive boundary conditions on the downstream perturbations. The upstream porosity is constant ($\Hp^u = 1$) and therefore the pressure field satisfies a Laplace equation. The upstream velocity perturbations are then:
\begin{equation}
\delta \Hv_\xi^u = A_v e^{\Hu\xi} \cos(\Hu\eta)e^{\Ho \tau}, \ \ \ \ \delta \Hv_\eta^u = - A_v e^{\Hu\xi}\sin(\Hu\eta)e^{\Ho \tau}, \label{eq:dvu}
\end{equation}
where $A_v$ is a constant that can be determined by matching to the downstream perturbation (Sec. \ref{sec:bcf})

The base concentration in the upstream region, from \eqref{cbase} with the right hand side set to zero, is
\begin{equation}
\Hc_b^u = 1 + A_c e^{\xi\Pek} \label{cbaseu},
\end{equation}
satisfying the boundary condition $\Hc_b^u(-\infty) = 1$. We can obtain the coefficients $A_c$ and $\Hc_b(0)$ by matching the convective and diffusive fluxes at the boundary between the upstream and downstream regions ($\xi = 0$):
\begin{align}
\Hc_b^u(0) &= \Hc_b(0), \\
(\Dx \Hc_b^u)_0 &= (\Dx \Hc_b)_0.
\end{align}
This gives the base concentration fields in Eqs. (9) and (10), which in dimensionless form are:
\begin{align}
\Hc_b^u &= 1 - \frac{e^{\xi\Pek}}{1+\Pek} && \xi < 0, \label{cbaseusol}\\
\Hc_b &= \frac{e^{-\xi}}{1+\Pek^{-1}} && \xi < 0. \label{cbasesol}
\end{align}

The upstream solution for the concentration perturbation follows from Eq.~\eqref{transport2} with $f_\phi=0$:
\begin{equation}
\left[\Pek^{-1}(\Dx^2 - \Hu^2) - \Dx \right]f_c^u = - \frac{A_v}{(1+\Pek^{-1})} e^{(\Pek+\Hu)\xi}.
\end{equation}
After eliminating the diverging (at $-\infty$) solution, the upstream concentration perturbation is,
\begin{equation}
f_c^u = A_c e^{(\Pek+\sqrt{\Pek^2+4\Hu^2})\xi/2} - \frac{A_v}{(1+\Pek^{-1})\Hu}e^{(\Pek+\Hu)\xi}; \label{eq:dcu}
\end{equation}
the constant $A_c$ is determined by matching the upstream and downstream perturbations to the reactant concentration and flux (see Sec. \ref{sec:bcf})

\section{Boundary conditions at the front}\label{sec:bcf}

The boundary condition on the downstream porosity, $\Hp(\xi_f ) = 1$, must be evaluated at the front $\xi=\xi_f $, and then referred to the mean position of the front $\xi=0$ by linearization,
\begin{equation}
\Hp(\xi_f ) = \Hp_b(0) + \xi_f  (\Dx \Hp_b)_0 + \delta \Hp(0). \label{eq:flin}
\end{equation}
The perturbation of the front, (relative to its mean position) is assumed to grow exponentially in time, along with the other fields,
\begin{equation}
\xi_f (\eta, \tau) = \xi_0\cos(\Hu \eta) e^{\Ho \tau}, \label{eq:bcf}
\end{equation}
with an initial amplitude $\xi_0$. The boundary condition on the downstream perturbation in porosity is then
\begin{equation}
f_\phi(0) = \xi_0(-\Dx \Hp_b)_0. \label{eq:bcphif}
\end{equation}

Continuity of velocity can be established at the mean front position directly, $\delta\Vv(0) = \delta\Vv^u(0)$, because the base velocity field is constant. Using Eq.~\eqref{eq:dv} for $\Hv_\xi$ and the incompressibility condition \eqref{cont2} for $\Hv_\eta$, the velocity field at the mean front position is
\begin{equation}
\Hv_\xi(0) = f_v(0)\cos(\Hu\eta)e^{\Ho \tau}, \ \ \ \ \Hv_\eta(0) = (\Dx f_v)_0\frac{-\sin(\Hu\eta)}{\Hu}e^{\Ho \tau}.
\end{equation}
Matching to the upstream velocity field \eqref{eq:dvu}, we obtain the boundary condition for the downstream velocity perturbation
\begin{equation}
(\Dx f_v)_0 - \Hu f_v(0) = 0. \label{eq:bcvf}
\end{equation}

Continuity of concentration and reactant flux requires two matching conditions at the front:
\begin{equation}\label{eq:bccf0}
\Hc(\xi_f ) = \Hc^u(\xi_f ), \ \ \ \ (\Dx \Hc)_{\xi_f } = (\Dx \Hc^u)_{\xi_f }.
\end{equation}
Expanding about the base profiles, as in Eq.~\eqref{eq:flin}, we obtain matching conditions on the downstream concentration perturbation:
\begin{align}
f_c(0) - f_c^u(0) &= 0, \label{eq:bccf1} \\
(\Dx f_c)_0 - (\Dx f_c^u)_0 &= -\xi_0\Pek, \label{eq:bccf2}
\end{align}
where we have again made use of the base solutions, \eqref{cbaseusol} and \eqref{cbasesol}.
The upstream perturbation, $f_c^u$ \eqref{eq:dcu}, can then be used to obtain a boundary condition on the downstream concentration perturbation,
\begin{equation}
(\Dx f_c)_0 + (\beta-\Pek)f_c(0) + \frac{\beta+\Hu}{(1+\Pek^{-1})\Hu}f_v(0)
= - \xi_0\Pek.\label{eq:bccf}
\end{equation}
Continuity of velocity, $f_v(0) = A_v$, was used to eliminate $A_v$, while Eq.~\eqref{eq:bccf1} was used to eliminate $A_c$ in favor of $f_c(0)$. The parameter $\beta$ is a function of $\Pek$ and $\Hu$,
\begin{equation}
\beta = \frac{1}{2}\left(\Pek-\sqrt{\Pek^2+4\Hu^2}\right). \label{eq:beta}
\end{equation}

Finally, the boundary conditions at the front, \eqref{eq:bcphif}, \eqref{eq:bcvf}, and \eqref{eq:bccf} can be expressed in terms of $f_\phi$ by using the relations $f_v = -e^\xi {\cal L}_\lambda(\partial_\xi - \Ho) f_\phi$ and $f_c = - (1+\Pek^{-1})^{-1}(\partial_\xi - \Ho) f_\phi$. Setting the amplitude of the perturbation $\xi_0 = 1$:
\begin{align}
f_\phi(0) &= 1, \label{eq:bcf1} \\
\left[(\Dx-\Hu+1){\cal L}_\lambda(\partial_\xi - \Ho) f_\phi\right]_0 &= 0, \label{eq:bcf2} \\
\left(1+\frac{\beta}{\Hu}\right) \left[{\cal L}_\lambda(\partial_\xi - \Ho) f_\phi\right]_0 + \left[(\Dx + \beta - \Pek)(\partial_\xi - \Ho)f_\phi\right]_0 &= 1+\Pek.\label{eq:bcf3}
\end{align}
Equation~\eqref{eq:pt} has been solved numerically by a spectral method \citep{Boyd1987}, which eliminates the diverging downstream solutions \eqref{eq:ptffsol} through the choice of basis functions. The system of equations for a given growth rate is closed by the first two conditions at the front \eqref{eq:bcf1}--\eqref{eq:bcf2}. The remaining boundary condition \eqref{eq:bcf3} is then used to determine the eigenvalue $\Ho$; the largest eigenvalue is used for the numerical results shown in Fig. 4. For further details on the spectral method, see  Sec. 5 of \cite{Szymczak2012}.

\section{Small porosity contrast: $\Delta \ll 1$}\label{sec:W=0}

In the limit that the porosity contrast $\Delta$ is small, we can make a regular perturbation expansion around $\Delta = 0$; to first order $W$ is independent of $\Hp$, $W = 3\Delta + {\cal O}(\Delta^2)$. Expanding $f_\phi$ and $\Ho$ in powers of $\Delta$,
\begin{equation}
f_\phi = f_0 + \Delta f_1 + \ldots, \ \ \ \ \Ho = \Ho_0 + \Delta \Ho_1 + \ldots, \
\end{equation}
there is a homogeneous equation for $f_0$, ${\cal L}_0 f_0 = 0$ \eqref{eq:pt}, with
\begin{equation}
{\cal L}_0 = (\partial_{\xi}^2 - \Hu^2)e^\xi{\cal L}_\lambda (\partial_\xi - \Ho_0).
\end{equation}
The zeroth-order solution satisfying the far-field boundary condition \eqref{eq:ptFF} is
\begin{equation}
f_0 = A_0e^{-(\Hu+1)\xi} + B_0e^{\lambda \xi},\label{eq:f0}
\end{equation}
where $\lambda$ is the negative root of the characteristic equation for ${\cal L}_\lambda$ \eqref{eq:Hlambda},
\begin{equation}
\lambda  = \frac{1}{2}\left(\Pek - \sqrt{(\Pek+2)^2+4\Hu^2}\right). \label{eq:lambda}
\end{equation}
The boundary condition \eqref{eq:bcf2} eliminates the first solution ({\it i.e.} $A_0 = 0$) and from \eqref{eq:bcf1} we have $f_0 = e^{\lambda\xi}$. The remaining boundary condition \eqref{eq:bcf3} gives an equation for the zeroth order growth rate
\begin{equation}
(\lambda + \beta - \Pek)(\lambda - \Ho_0) = 1 + \Pek,
\end{equation}
after noting that ${\cal L}_\lambda f_0 = 0$; the function $\beta(\Hu,\Pek)$ is given by Eq.~\eqref{eq:beta}. The solution,
\begin{equation}
\Ho_0 = \beta, \label{eq:w0}
\end{equation}
shows that in the absence of a permeability contrast the planar reaction front is stable ($\beta < 0$).

Taking the first-order (in $\Delta$) terms in Eq.~\eqref{eq:pt} we obtain the inhomogeneous equation, ${\cal L}_0 f_1 = 3\Hu^2 f_0$; the other first-order terms vanish because ${\cal L}_\lambda f_0 = 0$. Solving for $f_1$, again including only decaying solutions, we have
\begin{equation}
f_1 = A_1e^{-(\Hu+1)\xi} + B_1e^{\lambda \xi} + C_1 e^{(\lambda-1)\xi},\label{eq:f1}
\end{equation}
where $C_1$ is fixed by the zeroth order solution; by substitution,
\begin{equation}
C_1 = \frac{3\Hu^2\Pek}{(1 + \Pek - 2\lambda)(1 + \beta - \lambda)(\lambda+\Hu)(\lambda-\Hu)}.
\end{equation}

The boundary conditions at first-order in $\Delta$ are:
\begin{align}
f_1(0) &= 0, \label{eq:bc1} \\
\left[(\Dx-\Hu+1){\cal L}_\lambda(\partial_\xi - \beta) f_1\right]_0 &= 0, \label{eq:bc2} \\
\left[\left\{\left(1+\frac{\beta}{\Hu}\right) {\cal L}_\lambda + \Dx + \beta - \Pek\right\}(\partial_\xi - \beta)f_1\right]_0 &= (\lambda+\beta-\Pek)\Ho_1,\label{eq:bc3}
\end{align}
again making use of the relation ${\cal L}_\lambda f_0 = 0$ and $\Ho_0 = \beta$. Only $A_1$ and $C_1$ enter into the expression for the second boundary condition \eqref{eq:bc2}, so we can solve directly for $A_1$,
\begin{equation}
A_1 = \frac{3\Pek}{2(2+\Pek)(\lambda+\Hu)(1+\beta+\Hu)}.
\end{equation}
Next we use the first boundary condition \eqref{eq:bc1} to get $B_1$,
\begin{equation}
B_1 = -A_1 - C_1.
\end{equation}
Finally, the third boundary condition \eqref{eq:bc3} can be used to obtain an explicit expression for the first-order growth rate:
\begin{align}
\Ho_1 &= \frac{3(1+\Pek+2\Hu)(\beta+\Hu)}{2(1+\beta+\Hu)(\lambda+\Hu)(\lambda+\beta-\Pek)} \nonumber \\ &+\frac{3u\left[3+4\Pek+\Pek^2+(1+\Pek)\Hu+2\Hu^2-(3+\Pek+2\Hu)\lambda\right](\beta+\Hu)}{(1+\Pek-2\lambda)(1+\beta-\lambda)(1+\Pek+\Pek\lambda)(\lambda+\beta-Pe)}. \label{eq:w1}
\end{align}
We used Maple to check many of the calculations in this section; a worksheet {\it auxiliary.mw} is included in the Auxiliary Material.

\end{document}